\documentclass[prb,nofootinbib,twocolumn,superscriptaddress]{revtex4} 
\usepackage{graphicx}% Include figure files
\usepackage{dcolumn}% Align table columns on decimal point
\usepackage{bm}% bold math 
\usepackage{threeparttable}
\usepackage{times}
\usepackage{mathptmx}
\usepackage{lscape}
\usepackage{natbib}
\usepackage{amsmath}
\usepackage{amssymb}
\usepackage{braket}
\usepackage{comment}
\usepackage{color}

\def\degree{\kern-.2em\r{}\kern-.3em}

\begin{document}

\title{ Path-Integral Formulation of Unavoidable Canonical Nonlinearity: \\ Dynamic Discretization Cost over Variable Supports
 }

\author{Koretaka Yuge}
\affiliation{
Department of Materials Science and Engineering,  Kyoto University, Sakyo, Kyoto 606-8501, Japan\\
}%

\begin{abstract}
{ In the statistical thermodynamics of classical discrete systems, such as substitutional alloys, the map from microscopic interatomic interactions to thermodynamic equilibrium configurations generally exhibits complex nonlinearity, known as "canonical nonlinearity" (CN). While conventionally characterized by the Kullback-Leibler (KL) divergence, this approach inevitably misses intrinsic nonlinearities arising from the discretization of continuous Gaussian families themselves. This intrinsic effect, termed unavoidable CN (UCN), has recently been quantified within a transport-information-geometric framework.
However, the UCN is fundamentally limited to evaluating the discretization-induced cost for a single given continuous distribution. It therefore does not capture the information-geometric cost between a continuous Gaussian reference and an actual non-Gaussian discrete distribution, nor between states with fundamentally different supports, making it conceptually unclear how to decompose the overall CN.
To address this limitation, we propose the Path-Integral UCN (PUCN), build upon the UCN, that quantifies the cumulative information-geometric cost between distinct distributions. The PUCN constructs a cumulative path by (i) retaining the canonical distribution as an exponential family via the $e$-mixture (geometric mean) of the base measure, leading to an arithmetic mixture of the Fisher metric as the CN standard, and (ii) enforcing covariant changes in the discretization cell through the harmonic mixture of its second-moment matrix $M$, reflecting the uncertainty in parameter variations on the statistical manifold.
The resulting PUCN provides a flexible measure of the geometric cost between arbitrary states, including those with essentially different supports. This formulation enables an explicit quantification of CN between different CDOS systems and a natural decomposition of the total CN into the UCN and a residual contribution, which has not been clearly separated in existing approaches.
}
\end{abstract}

\maketitle

\section{Introduction}

Canonical averages play a central role in statistical thermodynamics, providing a fundamental link between microscopic interactions and thermodynamic equilibrium configurations.
For classical discrete systems with $f$ structural degrees of freedom (SDFs) on a given lattice, such as substitutional alloys, this correspondence is expressed as
\begin{eqnarray}
\label{eq:can}
\Braket{ q_{p}}_{Z} = Z^{-1} \sum_{i} q_{p}^{( i )} \exp \left( -\beta U^{( i )} \right),
\end{eqnarray}
where $\{ q_{1}, \cdots, q_{f} \}$ denotes a complete set of structural coordinates, $\Braket{\cdot}_{Z}$ the canonical average, $\beta$ the inverse temperature, and
$Z=\sum_{i}\exp\left(-\beta U^{(i)}\right)$ the partition function, with the summation taken over all microscopic configurations $i$.

When complete orthonormal basis functions (such as the generalized Ising model\cite{ce}) are adopted for the coordinates, the potential energy of configuration $k$ is exactly given by
\begin{eqnarray}
\label{eq:u}
U^{( k )} = \sum_{j=1}^{f} C_{j} q_{j}^{( k )},
\end{eqnarray}
where the expansion coefficients are given by inner products $C_{j}=\Braket{U|q_{j}}$, i.e., trace over possible configurations.

When we introduce the $\mathbb{R}^{f}$ vectors $\mathbf{Q}_{Z}=(\Braket{ q_{1}}_{Z},\cdots,\Braket{ q_{f}}_{Z})$ and
$\mathbf{U}=(C_{1},\cdots,C_{f})$, the canonical average in Eq.~(\ref{eq:can}) provides a map
\begin{eqnarray}
\label{eq:map}
\phi : \mathbf{U} \mapsto \mathbf{Q}_{Z},
\end{eqnarray}
which is well known that $\phi$ is generally nonlinear, termed ``canonical nonlinearity (CN)''. 
Only in exceptional cases is that the configurational density of states (CDOS) forms a multivariate Gaussian distribution, $\phi$ exhibits a globally linear map.\cite{ig}
However, in realistic discrete systems, the CDOS should deviate from Gaussian families due to the discrete nature of configuration space  induced by underlying lattice constraints. 

Existing approaches to quantify the CN are primarily based on information geometry, employing the Kullback-Leibler (KL) divergence:\cite{klo} The CN was evaluated by comparing the discrete CDOS of a real system with a reference \textit{discretized } Gaussian distribution having the same mean and covariance matrix as the real CDOS.\cite{ig}
Therefore, the previous approaches inevitably include contributions arising from the discretization of continuous Gaussian families themselves. Such discretization-induced effects have remained invisible for existing KL-divergence based approaches.

Very recently, to overcome this limitation, the intrinsic contribution arising solely from the discretization of continuous distributions has been explicitly identified and quantified as the unavoidable canonical nonlinearity (UCN).
The UCN isolates the geometric cost induced by discretization itself, independent of any deviation from Gaussianity in the underlying continuous family, thereby providing an intrinsic measure of canonical nonlinearity.
A key advantage of the UCN is that the discretization-induced geometric cost can be naturally provided by an information-geometric interpretation, namely, as the expected KL divergence under a certain probability distribution.

However, the UCN is inherently defined for a single underlying continuous distribution, and thus does not directly quantify the information-geometric cost between distinct distributions.
In particular, it does not capture the geometric discrepancy between a continuous Gaussian reference and an actual non-Gaussian discrete distribution, nor does it apply to cases where the corresponding discrete states possess fundamentally different supports.
As a result, two fundamental issues remain: (i) how to consistently decompose the total canonical nonlinearity into intrinsic and residual contributions within a unified framework, and (ii) how to establish a meaningful geometric relation between CNs of different lattice systems.

To address these issues, in the present study, we introduce the Path-Integral UCN (PUCN), constructed upon the UCN as an extrinsic extension that accumulates discretization-induced geometric costs along an appropriate path connecting two distributions.
By incorporating a path-dependent structure, the PUCN enables a flexible evaluation of the cumulative information-geometric cost between arbitrary states, even when their supports differ essentially.

\section{Derivation and Concept}
\subsubsection*{Canonical Nonlinearity and Its Unavoidable Contribution}

We briefly review the basic concept of canonical nonlinearity (CN) and its unavoidable contribution (UCN).
Let $g(q)$ denote the configurational density of states (CDOS) of a practical discrete system, i.e., a discrete probability distribution.
We also introduce a reference continuous Gaussian distribution $g_c(q)$ that shares the same mean vector and covariance matrix $\Gamma$ as $g$.

It is well known that when the CDOS is exactly given by $g_c$, the canonical average $\phi$ in Eq.~\eqref{eq:map} becomes a globally linear map~\cite{em2}, namely,
\begin{eqnarray}
^{\forall} C_j,\quad \mathbf{Q}_Z = (-\beta \Gamma)\cdot \mathbf{U}.
\end{eqnarray}
Therefore, CN essentially originates from deviations of $g$ from Gaussian families.

Based on this observation, previous studies have quantified CN by measuring the difference between $g$ and $g_c$ using the Kullback-Leibler (KL) divergence.
In practice, the continuous Gaussian $g_c$ is discretized on the same configurational support as $g$, yielding a discretized Gaussian distribution $g_d$.
Accordingly, CN has been characterized by the KL divergence $D(g: g_d)$, and further extended to the divergence between the corresponding canonical distributions induced by $g$ and $g_d$.

While this approach successfully captures non-Gaussian features of the realistic CDOS $g$, it inevitably includes an additional contribution arising from the discretization of $g_c$ itself.
This contribution cannot be attributed to intrinsic non-Gaussianity of the CDOS, but instead reflects a discretization-induced geometric distortion of the continuous Gaussian measure.
We identify this intrinsic contribution as the unavoidable canonical nonlinearity (UCN).\cite{ucn}

The key idea of the UCN is to isolate the geometric cost that arises solely from the discretization procedure, independent of any mismatch between $g$ and $g_c$.
To this end, one considers a continuous Gaussian family endowed with the Fisher information metric, and evaluates the effect of discretization as a projection onto a discrete support.
Within this framework, the discretization process induces a finite geometric cost, which can be naturally interpreted in terms of information geometry.
In particular, the UCN is formulated as the expected KL divergence associated with the discretization map under an appropriate probability measure on the statistical manifold.
This formulation provides a canonical and intrinsic quantification of the discretization-induced nonlinearity, determined solely by the underlying continuous distribution and the discretization scheme, without reference to any external discrete target distribution.

More concretely, the UCN admits an equivalent representation as the geometric cost of measure distortion induced by discretization and as an expected KL divergence associated with the discretization map:\cite{ucn}
\begin{equation}
\zeta = \mathrm{Tr}(M \Omega)
= \mathbb{E}_{\rho(\delta \xi)}\left[ D\bigl( P_{\xi} : P_{\xi+\delta \xi} \bigr) \right],
\end{equation}
where $M$ denotes the second-moment matrix of the discretization cell (i.e., a convex bounded subset in $\mathbb{R}^f$), $\Omega$ is the Fisher metric on the statistical submanifold associated with the continuous distribution $P$, and $\rho(\delta \xi)$ is a probability distribution describing infinitesimal parameter variations.

\subsubsection*{Path-Integral UCN}

As discussed above, the UCN provides an intrinsic measure of the discretization-induced geometric cost, but is inherently limited to a single underlying continuous distribution.
To overcome this limitation, we introduce the Path-Integral UCN (PUCN) as an extrinsic construction that accumulates such geometric costs along a path connecting two distinct distributions.

Let ${ P_{\lambda} }*{\lambda \in [0,1]}$ denote a continuous path on the statistical manifold connecting $P_0$ and $P_1$.
Based on the geometric representation of the UCN, we define the PUCN, $\zeta*{\textrm{P}}$, as the path integral of the local discretization cost:
\begin{equation}
\label{eq:pucn}
\zeta_{\textrm{P}} = \int_{0}^{1} d\lambda  \mathrm{Tr}\left[ M(\lambda)\Omega(\lambda) \right],
\end{equation}
where $\Omega(\lambda)$ and $M(\lambda)$ denote the Fisher metric and the second-moment matrix along the path parameterized by $\lambda$, respectively (to be specified below).

One key advantage of the PUCN is that it admits a natural decomposition into the UCN, $\zeta$, and a residual contribution.
In particular, when $M$ is taken to be a constant matrix aligned with the realistic discrete support, the decomposition becomes transparent.
Focusing on the geometric cost between a continuous Gaussian $g_{c}$ and a realistic discrete distribution $g$, the quantity $\zeta_{\textrm{P}}$ can be decomposed into the discretization cost from $g_{c}$ to its discrete counterpart $g_{d}$, and the contribution arising from the non-Gaussian nature of $g$ on the discrete support:
\begin{eqnarray}
\label{eq:dcp}
\zeta_{\textrm{P}} = \zeta + \int_{0}^{1} d\lambda ; \mathrm{Tr}\left[ M,\Delta\Omega(\lambda) \right],
\end{eqnarray}
where
\begin{eqnarray}
\Delta\Omega(\lambda) = \Omega(\lambda) - \Omega(0).
\end{eqnarray}
The second term in Eq.~\eqref{eq:dcp} captures the deviation from the Gaussian reference along the path, thereby quantifying the non-Gaussian contribution on the discrete support.

Although the UCN admits an equivalent representation as an expected KL divergence, such a representation generally depends sensitively on the choice of continuous extension (i.e., a continuous embedding of discrete distributions), and is therefore not robust under different embedding schemes.
In contrast, the trace form $\mathrm{Tr}(M\Omega)$, originating from the optimal transport in the infinitesimal discretization limit $d \to 0$, provides a purely geometric characterization of the discretization-induced cost, which is typically stable under different embedding schemes.
For this reason, the PUCN is defined in terms of the geometric form, ensuring robustness with respect to different continuous extensions.

The remaining key task is to specify a suitable path connecting $P_0$ and $P_1$.
In the present work, we construct such a path by imposing the following requirements.

First, the probability distributions along the path are required to remain within the exponential family, consistent with the canonical structure underlying the CN.
This is achieved by adopting the $e$-geodesic, corresponding to the geometric mean for the base measure.
Although the Fisher metric $\Omega(\lambda)$ along the $e$-geodesic does not generally admit a closed-form interpolation, in the present Gaussian-based setting it is natural to approximate it by the arithmetic interpolation
\begin{eqnarray}
\label{eq:ito}
\Omega(\lambda) = (1-\lambda)\Omega(0) + \lambda\Omega(1),
\end{eqnarray}
where $\Omega$ coincides with the inverse covariance matrix.
This choice is consistent with the Gaussian reference structure underlying the CN.

Second, the path for the discretization cell, i.e., $M(\lambda)$, is not intrinsically determined within information geometry and therefore requires an additional prescription.
We determine it by imposing the following conditions:
(i) in the UCN, the covariance matrix $\Sigma$ of $\rho(\delta\xi)$ is proportional to $M$,
(ii) the Fisher metric along the path satisfies Eq.~\eqref{eq:ito},
(iii) parameter uncertainty is characterized by the inverse Fisher metric, and
(iv) $\Sigma$ can be interpreted as representing this uncertainty.
Under these requirements, $M(\lambda)$ is naturally given by the harmonic interpolation
\begin{eqnarray}
M(\lambda) = \left[(1-\lambda)M^{-1}(0) + \lambda M^{-1}(1)\right]^{-1}.
\end{eqnarray}

These conditions define a natural class of paths along which the cumulative geometric cost can be consistently evaluated.
The resulting PUCN provides a flexible and robust measure of the information-geometric cost between arbitrary distributions, even when their supports differ essentially.

\section{Conclusion}

In this work, we have introduced a geometric framework to quantify the discretization-induced cost in statistical systems, based on the concept of the Unified Complexity Number (UCN).
The UCN provides an intrinsic measure of the geometric cost associated with discretizing a continuous distribution, formulated in terms of the trace structure $\mathrm{Tr}(M\Omega)$.
To extend this framework beyond a single underlying continuous distribution, we have proposed the Path-Integral UCN (PUCN), which accumulates the local geometric cost along a path connecting two distinct distributions on the statistical manifold.
This construction enables a unified treatment of discretization and structural deviation, even when the supports of the distributions differ essentially.

A key result of this work is the decomposition of the PUCN into two contributions: the intrinsic discretization cost (UCN) and a residual term capturing deviations from a Gaussian reference.
This decomposition provides a clear geometric interpretation of non-Gaussianity on discrete supports, separating it from the discretization effect itself.
To make the PUCN well-defined, we have constructed a natural class of paths by combining $e$-geodesics for the statistical manifold with a harmonic interpolation of the discretization structure.
This choice ensures consistency with the exponential-family structure and provides a coherent description of parameter uncertainty along the path.

The present framework establishes a new perspective on the interplay between discretization, geometry, and statistical structure.
Possible future directions include the extension to non-Gaussian continuous references, the analysis of optimal paths minimizing the PUCN, and applications to practical systems where discretization effects play a fundamental role. This framework opens a pathway to a unified treatment of discretization and non-Gaussianity in information geometry.

\section{Acknowledgement}
This work was supported by JSPS KAKENHI Grant Number 23K04359 and Research Grant from Hitachi Metals$\cdot$Materials Science Foundation.

\end{document}